\documentclass[twocolumn,aps,amssymb,floatfix]{revtex4}

\usepackage{graphicx}
\usepackage{bm}
\usepackage{amsmath}
\usepackage{amssymb}
\usepackage{amsfonts}
\usepackage{float}
\usepackage{hyperref}
\usepackage{dsfont}  
\usepackage{slashed}  
\usepackage{booktabs}
\usepackage{multirow}
\usepackage{subfigure}
\usepackage[sort&compress]{natbib}

\newcommand{\be}{\begin{equation}}  
\newcommand{\ee}{\end{equation}}  
\newcommand{\beq}{\begin{eqnarray}}  
\newcommand{\eeq}{\end{eqnarray}}

\newcommand{\Op}{\mathcal{O}}

\begin{document}
\title{Direct Evaluation of the Quark Content of  Nucleons from Lattice QCD at
  the Physical Point}
\author{
  A.~Abdel-Rehim$^{1}$,
  C.~Alexandrou$^{1,2}$,
  M.~Constantinou$^{1}$,
  K.~Hadjiyiannakou$^{1,2}$,
  K.~Jansen$^{3}$,
  Ch.~Kallidonis$^{1}$,
  G.~Koutsou$^{1}$,
  A.~Vaquero Avil\'es-Casco$^{4}$
}
\affiliation{
  $^1$Computation-based Science and Technology Research Center,
  The Cyprus Institute,
  20 Kavafi Street,
  Nicosia 2121,
  Cyprus \\
  $^2$Department of Physics,
  University of Cyprus,
  P.O. Box 20537,
  1678 Nicosia,
  Cyprus\\  
  $^3$NIC, DESY,
  Platanenallee 6,
  D-15738 Zeuthen,
  Germany\\
  $^{4}$ INFN Sezione di Milano-Bicocca, Edificio U2, Piazza della
  Scienza 3, 20126 Milano, Italy
}

\begin{abstract}
  We evaluate the light, strange and charm scalar content of the
  nucleon using one lattice QCD ensemble generated with two degenerate light
  quarks with mass fixed to their physical value. We use improved
  techniques to evaluate the disconnected quark loops to sufficient
  accuracy to determine the strange and charm nucleon $\sigma$-terms
  in addition to the light quark content $\sigma_{\pi N}$.
We find $\sigma_{\pi
    N}$=$37.2(2.6)(^{4.7}_{2.9})$~MeV,
  $\sigma_{s}$=$41.1(8.2)(^{7.8}_{5.8})$~MeV and
  $\sigma_c$=$79(21)(^{12}_{\phantom{1}8})$~MeV, where the first error is
  statistical and the second is  the systematic error due
  to the determination of the lattice spacing, the assessment of
  finite volume and residual excited state effects.
\end{abstract}
\maketitle 
\bibliographystyle{apsrev}

\textit{Introduction:} The scalar quark contents of the nucleon, or the
so-called nucleon $\sigma$-terms, are fundamental quantities of QCD
determining the mass generated by the quarks in the nucleon and thus
related to the explicit breaking of chiral symmetry. They are relevant
for a wide range of physical processes, such as pion- and kaon-nucleon
scattering, but also for the interpretation of direct-detection dark
matter (DM) searches being undertaken by a number of
experiments~\cite{Cushman:2013zza}. DM candidates under consideration
are weakly interacting massive particles (WIMPs) predicted in many
beyond the standard model theories that interact with normal matter by
elastic scattering with nuclei~\cite{Bertone:2004pz}. In such a
process, a WIMP, due to its large mass, produces a Higgs boson that
interacts with the nucleon via scalar density operators. At zero
momentum transfer, the cross section for spin independent elastic
scattering depends quadratically on the nucleon scalar matrix element,
which constitutes the largest uncertainty~\cite{Ellis:2008hf}. It is
customary to define the nucleon $\sigma$-terms to be scheme- and
scale-independent quantities:
\be
\sigma_f=m_{q_f} \langle N|\bar{q}_f q_f|N\rangle ,\,\,
\sigma_{\pi N}= m_{ud}\langle N | \bar{u}u + \bar{d}d | N \rangle
\ee
for a given quark $q_f$ of flavor $f$,  or for the isoscalar
combination, where $m_{q_f}$ is the mass of $q_f$,
$m_{ud}=(m_u+m_d)/2$ is the average light quark mass and $|N\rangle$
is the nucleon state.
 Since the pioneering chiral perturbation theory analysis that yielded
$\sigma_{\pi N}\sim 45$~MeV~\cite{Gasser:1990ce}, there has been significant progress in the
determination of $\sigma_{\pi N}$ from experimental data~\cite{Hoferichter:2012wf,Alarcon:2011zs}. Using
high-precision data from pionic atoms to determine the $\pi
N$-scattering lengths and a system of Roy-Steiner equations that
encode constraints from analyticity, unitarity, and crossing symmetry
a value of $59.1(3.5)$~MeV is obtained~\cite{Hoferichter:2015dsa}.
This larger value of $\sigma_{\pi N}$ has theoretical implications on
our understanding of the strong interactions as stressed in
Ref.~\cite{Leutwyler:2015jga}. The determination of the strange
$\sigma_s$-term is more difficult since it would require an analysis
of kaon-nucleon scattering phase shifts. Alternatively, one can use as
input the values of $\sigma_{\pi N}$ and of the ratio $m_s/m_{ud}$,
and SU(3) chiral perturbation theory to determine it.
The evaluation of the charm scalar content is even harder amplifying
even further the uncertainty in the WIMP-nucleon cross-section.

Given the importance of these quantities, a number of lattice QCD
calculations have been undertaken using two
approaches~\cite{Young:2009ps}. The first uses the Feynman-Hellmann
theorem that is based on the variation of the nucleon mass $m_N$ with
$m_{q_f}$:~$\sigma_{f}=m_{q_f}\frac{\partial m_N}{\partial  m_{q_f}}$.
Within lattice QCD the mass of the nucleon has been computed using
various discretized actions and different values of the light quark
mass including recent simulations with the physical value (referred to
as the physical point).
The most extensive analysis, using a large set of simulations with two light and a strange clover-improved fermions  reaching physical values of the pion mass and including an
assessment of finite volume effects and an extrapolating to the
continuum limit, yielded a precise value of $\sigma_{\pi
  N}=38(3)(3)$~MeV~\cite{Durr:2015dna}.
However, using the same approach to evaluate $\sigma_s$ yields larger
errors. The analysis of  Ref.~\cite{Durr:2015dna} finds
$\sigma_s=105(41)(37)$~MeV, while for $\sigma_c$ this approach is
currently not applicable.
   
An alternative method is to evaluate directly the nucleon matrix
elements of the scalar operator that involves 
disconnected quark loops. These are  much more demanding to compute
than hadron masses. Therefore, it is only recently that
a direct computation of the $\sigma$-terms has been performed using
dynamical simulations~\cite{Bali:2011ks,Freeman:2012ry,Gong:2013vja,Abdel-Rehim:2013wlz,Alexandrou:2013nda,Yang:2015uis}.
In this Letter we employ improved methods to compute the disconnected quark
loops, and evaluate $\sigma_{\pi N}$, $\sigma_s$ and $\sigma_c$
directly at the physical point, thus eliminating chiral fits. The
evaluation is done for one ensemble of $N_f$=2 twisted mass fermions
at maximal twist with a lattice spacing of $a$=0.093(1)~fm determined
from the nucleon mass~\cite{Abdel-Rehim:2015owa}. An analysis
using a single ensemble precludes the evaluation of cut-off and
infinite-volume effects using directly lattice QCD results  on these
quantities. However, we present  
conservative estimates of these systematics using auxiliary
arguments. 

Since we are using an $N_f$=2 ensemble, the strange and charm quarks
are absent from the sea. Results obtained using $N_f$=2 and
$N_f$=2+1+1 twisted mass ensembles on quantities for which such
quenching can potentially have a large effect, such as the strange and
charm quark masses~\cite{Blossier:2010cr,Carrasco:2014cwa} or the kaon
and D-meson decay constants~\cite{Blossier:2009bx,Carrasco:2014poa},
showed no detectable quenching effects.
For the value of $\sigma_s$ itself, there are two direct computations of the matrix element
 using the same overlap fermion ensembles, one with $N_f$=2~\cite{Takeda:2010cw} and one with $N_f$=2+1~\cite{Oksuzian:2012rzb} both yielding consistent values albeit with a large systematic uncertainty.
A more recent direct evaluation of the matrix element with  overlap
valence quarks on $N_f$=2+1 domain wall sea fermions including the physical point  yielded a value  $\sigma_s=32.3(4.7)(4.9)$, which is  compatible with the value of $\sigma_s=30(8)(21)$~MeV obtained  with $N_f=2$ overlap fermions~\cite{Takeda:2010cw}.
These results suggest that quenching effects are negligible within the statistical accuracy that they are computed presently. We, therefore, neglect quenching effects in this Letter.


\textit{Matrix elements in lattice QCD:} The nucleon scalar matrix
element can be extracted from the ratio of the three-point function of
the scalar operator to the two-point function at zero momentum
transfer,
$R(t_s;t_\textrm{ins})=\frac{C_\textrm{3pt}(t_s-t_0,t_\textrm{ins}-t_0)}{C_\textrm{2pt}(t_s-t_0)}$~\cite{Abdel-Rehim:2015owa}.
Inserting a complete set of states yields
\begin{align}
  R(t_s, t_{\rm ins})=& 
  \frac{\sum\limits_{i,j=0}^{\infty}A_{ij}e^{-\delta E_i(t_s-t_\textrm{ins})}e^{-\delta E_j(t_\textrm{ins}-t_0)}}{1+\sum\limits_{i=1}^{\infty}c_ie^{-\delta E_i(t_s-t_0)}}\nonumber\\
  \xrightarrow[\delta E_i (t_s-t_{\rm ins})\gg 1]{\delta E_i(t_{\rm ins}-t_0)\gg 1}& \,m_{q_f}\langle N | \bar{q}_f q_f | N \rangle,\label{eq:ratio}
\end{align}
where $\delta E_i=E_i-E_0$ is the energy gap between the $i^{\rm th}$
nucleon excited state $E_i$ and the ground state $E_0=m_N$,
$A_{00}$ is the desired matrix element, $\Op=m_{q_f}\bar{q}_f q_f$,
and we consider the light, the strange and the charm quark flavors.
In Eq.~(\ref{eq:ratio}) the desired matrix element is obtained when
the insertion-source, $t_{\rm ins}-t_0$, and the sink-insertion,
$t_s-t_{\rm ins}$ time separations are large enough so that
contributions to the matrix element of $\Op$ from excited states with
the same quantum numbers as the nucleon are negligible. For the scalar
operator these have been shown to be large~\cite{Alexandrou:2013nda}.
We use three approaches in order to check that indeed excited state
contributions are sufficiently suppressed: i) In the so-called
\textit{plateau} method we look for the range of the values of  $t_{\rm ins}$
 for which  the ratio of Eq.~(\ref{eq:ratio}) becomes
time-independent (plateau region) and then fit a constant within this
region. This is done for several values of $t_s$. Excited states are
sufficiently suppressed when the value of the plateau remains
statistically unchanged. Smearing techniques are crucial to reduce the
coefficients $A_{ij}$ for $i,j>0$ and $c_i$ in the three- and
two-point functions. We use both gaussian and APE smearing to maximize
the overlap of our interpolating field with the
nucleon~\cite{Abdel-Rehim:2015owa}. ii) In the \textit{summation}
method~\cite{Martinelli:1987zd} we consider the ratio of
Eq.~(\ref{eq:ratio}) summed over $t_{\rm ins}$:
\begin{align}
  &R^{\rm sum}(t_s)=\sum_{t_{\rm ins}/a=1}^{t_s/a-1}R(t_s, t_{\rm ins})=&\nonumber\\
&\frac{
  \sum\limits_{i=0}^{\infty}A_{ii}(t_s/a-2)e^{-\delta E_it_s} + \sum\limits_{i\ne j=0}^{\infty}A_{ij}\frac{e^{-a\delta E_{ij}}-e^{-\delta E_{ij}t_s}
  }{
    1-e^{-a\delta E_{ij}}}e^{-\delta E_it_s}}{1+\sum\limits_{i=1}^{\infty}c_ie^{-\delta E_it_s}}&
  \label{eq:summed ratio}
\end{align}
where we set $t_0=0$ for
simplicity in what follows and define $\delta E_{ij}=E_j-E_i$. For large enough $t_s$, $R^{\rm sum}$
depends linearly on $t_s/a$, and the matrix element is given by $A_{00}$,
obtained as the slope of a linear fit. While excited state
contributions are suppressed as $e^{-\delta E_1 t_s}$, as compared to
$e^{-\delta E_1(t_s-t_{\rm ins})}$ and $e^{-\delta E_1t_{\rm ins}}$ in
the plateau method, fitting to a two-parameter linear dependence
increases the statistical errors.
iii) In the so-called \textit{two-state fit}, one takes into account
the contributions of the first excited state by fitting to the form
given in Eq.~(\ref{eq:ratio}) neglecting higher order terms. This
introduces five fit parameters, namely $A_{00}, A_{01}, A_{11}, c_1$
and $\delta E_1$. We consider that excited states are sufficiently
suppressed when the value extracted from the plateau method is
consistent between two $t_s$, as well as with the other two
approaches.

\textit{Computation of the three-point function:} The three-point
function $C_{\rm 3pt}(t_s, t_{\rm ins})$ receives contributions from
two types of diagrams, one when the scalar operator couples to a
valence quark inside the nucleon (connected) and one when it couples
to a sea quark (disconnected).
For $\sigma_{\pi N}$ both contributions are non-zero, while for
$\sigma_s$ and $\sigma_c$ only the disconnected contribution is
present. 
Evaluation of quark-disconnected contributions is notoriously difficult and
it is only very recently that such contributions can be computed
~\cite{Babich:2010at,Bali:2011ks,Oksuzian:2012rzb,Gong:2013vja,Alexandrou:2013nda,Abdel-Rehim:2013wlz,Yang:2015uis}.
This is due to the fact that one needs to evaluate a closed quark loop
of the form $\sum_{\vec{x}_{\rm ins}}{\rm Tr}[G_f(x_{\rm ins};x_{\rm
    ins})]$
where  $G_f(x;y)$ is the quark propagator.
Due to the appearance of the sum over the spatial coordinate
$\vec{x}_{\rm ins}$, the evaluation of disconnected contributions
requires knowledge of the quark propagator from all to all spatial
coordinates, which translates into spatial volume $N_S^3$ inversions
of the Dirac matrix compared to two per quark flavor required for the
evaluation of the connected contribution. This is overcome by using
stochastic techniques to evaluate the quark propagator entering the
loop.

We use the twisted mass fermion discretization
scheme~\cite{Frezzotti:2000nk}, which, besides ensuring ${\cal O}(a)$
improvement for physical observables after tuning to maximal
twist~\cite{Frezzotti:2003ni}, is particularly suited for evaluating scalar matrix elements. The first
important advantage is that all scalar matrix elements are
multiplicatively renormalizable~\cite{Dinter:2012tt} avoiding the
mixing that occurs between the bare light and strange scalar quark
matrix elements in other Wilson-type fermion discretizations. This
property also holds for chiral invariant lattice formulations, which
however are computationally much more demanding. The second important
advantage has to do with the twisted mass term of the doublets in the
action, which helps reduce the gauge noise of disconnected quark
loops. This is because the isoscalar combination of a flavor
doublet of the scalar operator transforms into an isovector of the
pseudo-scalar operator in the twisted mass formulation at maximal
twist. For the u- and d-flavor doublet we have
 $     \bar{u}u + \bar{d}d= i\bar{\chi}_{u}\gamma_5\chi_u - i\bar{\chi}_d\gamma_5\chi_d$,
where $\chi_u$ and $\chi_d$ are the two degenerate light quark fields
in the twisted mass basis. The disconnected quark loop contribution to $\sigma_{\pi
  N}$ therefore becomes~\cite{Michael:2007vn}: 
\beq
& & \sum_{\vec{x}_{\rm ins}}{\rm Tr}[i\gamma_5G_{\chi_u}(x_{\rm ins};x_{\rm ins})-i\gamma_5G_{\chi_d}(x_{\rm ins};x_{\rm ins})]\nonumber \\
& =&2\mu_l\sum_{y,\vec{x}_{\rm ins}}{\rm Tr}[\gamma_5G_{\chi_u}(x_{\rm ins};y)\gamma_5G_{\chi_d}(y;x_{\rm ins})].
\label{mu-reduction}
\eeq
The appearance of the small twisted light quark mass parameter $\mu_l$ allows for
significant reduction in the gauge noise. In this form, stochastic
techniques can be employed to obtain the trace via the so-called
\textit{one-end trick}~\cite{McNeile:2006bz} enabling the accurate
computation of the quark loops at all time insertions $t_{\rm
  ins}$~\cite{Alexandrou:2013wca,Abdel-Rehim:2013wlz}. Having the
quark loop for all $t_{\rm ins}$, the summation method can be employed
without any further cost, which is an additional advantage of this
formulation.

For the strange and the charm quarks a similar procedure can be
followed. In this case we consider Osterwalder-Seiler doublets~\cite{Frezzotti:2004wz}
to construct, in the twisted mass basis, the pseudo-scalar current:
$\frac{1}{2}(i\bar{\chi}_{f+}\gamma_5{\chi}_{f+}-i\bar{\chi}_{f-}\gamma_5{\chi}_{f-})$,
where $f=s,c$ and $f^{\pm}$ refers to taking $\pm \mu_f$.  The nucleon
matrix elements of these operators give $\sigma_s$ and
$\sigma_c$. Unlike $\sigma_{\pi N}$, however, only purely disconnected
contributions are involved.

\textit{Simulation parameters:} We use configurations simulated with
the $N_f$=2 twisted mass fermion action~\cite{Jansen:2009xp} including
a clover term with $c_{SW}=1.57551$ 
with $\mu_l$ tuned to obtained  the physical value
of the pion
mass~\cite{Abdel-Rehim:2015pwa,Abdel-Rehim:2014nka,Abdel-Rehim:2013wba,Abdel-Rehim:2013yaa,Deuzeman:2013xaa}. The
lattice size is $N_S=48$ in the spatial and $N_T= 96$ in the temporal
direction.
For the strange and charm doublets of Osterwalder-Seiler quarks, we
tune the bare strange and charm quark twisted mass $a\mu_s$ and
$a\mu_c$ to reproduce the experimental value of the $\Omega^-$
and $\Lambda_c$  mass, respectively, obtaining
$a\mu_s=$0.0264(3) and $a\mu_c=0.3348(15)$.

The statistical uncertainty in the disconnected quantities is reduced
as compared to that of Ref.~\cite{Abdel-Rehim:2013wlz} by increasing the
statistics at reduced cost.  This is accomplished by using a
polynomially accelerated implicitly restarted Arnoldi method to obtain
the $N_{ev}$ smallest eigenmodes of our linear system, which we use to
precondition the conjugate gradient (CG) algorithm for successive
solves on the same configuration. 
We take $N_{ev}=1400$, which yields a speed-up of about twenty
times as compared to the non-deflated CG.
In the stochastic evaluation of the quark
loops we use $Z(4)$ noise vectors with 
the number given in Table~\ref{table:statistics}. For the strange and
charm quark loops, we use the \textit{truncated solver method} (TSM),
where we combine a large number of stochastic estimates computed to
low precision with a small number of stochastic estimates computed to
high precision, appropriately
tuned~\cite{Bali:2009hu,Alexandrou:2012zz}. Table~\ref{table:statistics}
gives a summary of the statistics. All quark loop
contributions are evaluated using graphics cards where the QUDA
software is extended to include these
quantities~\cite{Alexandrou:2014fva}.   

\begin{table}[!h]
  \caption{$N_{\rm cnf}$ and  $N_{\rm src}$ is the number of
    configurations and source positions per configuration.
    For the disconnected loops, we give, in addition, the number of stochastic
    sources $N_r$, the number of high ($N_r^{hp}$)
    and low precision ($N_r^{lp}$) solves used within the TSM.    For the two-point functions we used $N_{\rm cnf}=1800$ and $N_{\rm src}=100$. 
}
  \label{table:statistics}
  \begin{tabular}{ccccccc}
    \hline\hline
    \multicolumn{3}{c}{Connected: three-point}        &
    \multicolumn{4}{c}{Fermion loop}        \\
    $t_s/a$ & $N_{\rm cnf}$ & $N_{\rm src}$ & Flav. & $N_\textrm{cnf}$ &  $N^{hp}_{r}$ & $N^{lp}_{r}$ \\
    \hline
    10,12,14&  192         & 16 & light   & 1800         & 2250 &0               \\
    16      &  265         & 88 & strange & 1800         & 63 & 1024              \\
    18      &  517         & 88 & charm   & 1800         & 5 & 1250             \\
     \hline\hline
  \end{tabular}
\end{table}

\begin{figure}[h!]
  \includegraphics[width=0.9\linewidth]{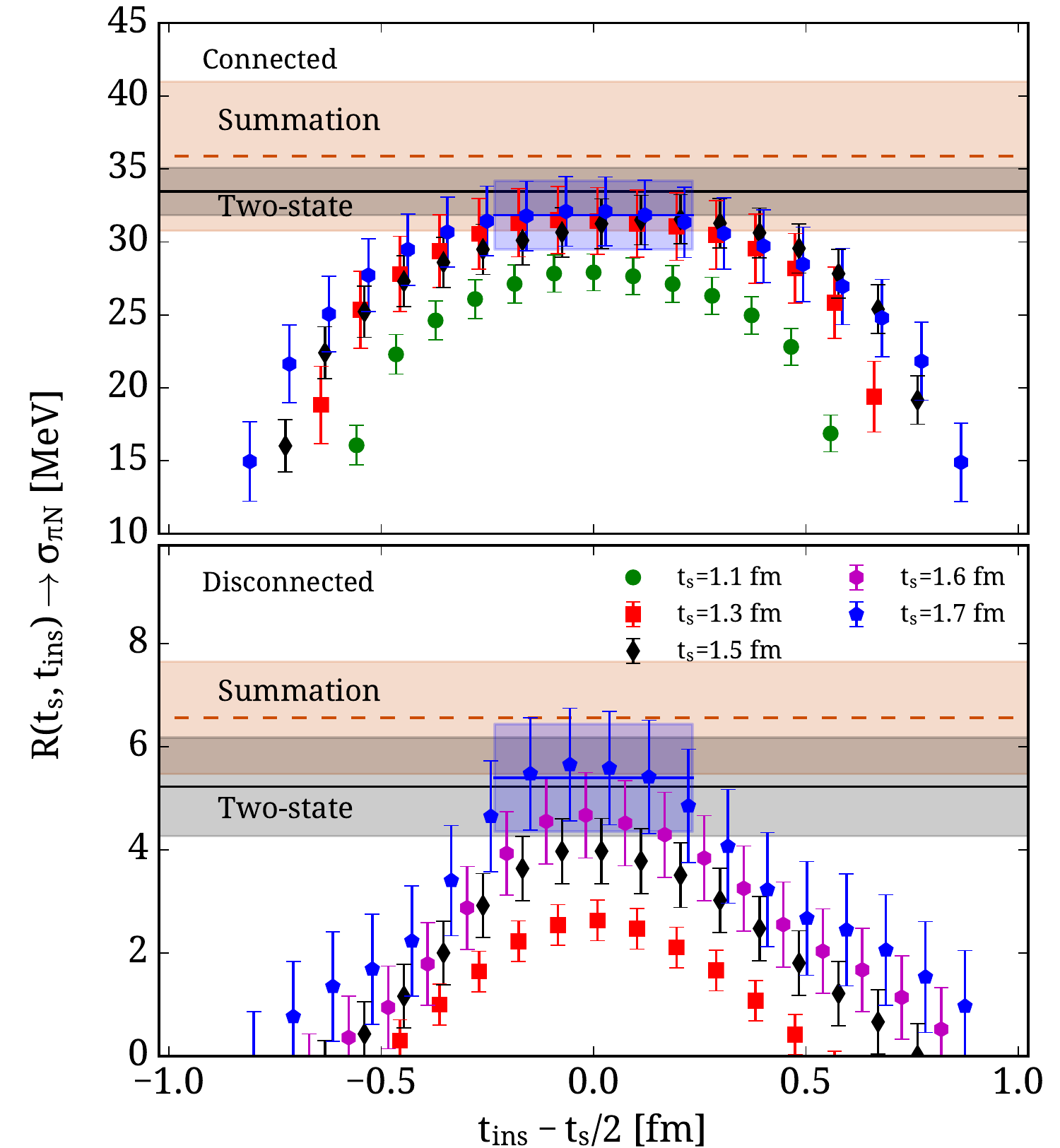}
  \caption{Ratios  yielding $\sigma_{\pi N}$, connected contribution
    (upper) and disconnected (lower), versus $t_{\rm ins}$ shifted by
    half the sink-source time separation. 
The lines and associated  bands show fits to the ratio in the plateau region (blue starting at the lower and finishing at the upper fit range used), the two-state fit (grey solid line) and
    the summation method (brown dashed line).}
  \label{fig:plateau light}
\end{figure}
\begin{figure}[h!]
  \includegraphics[width=0.9\linewidth]{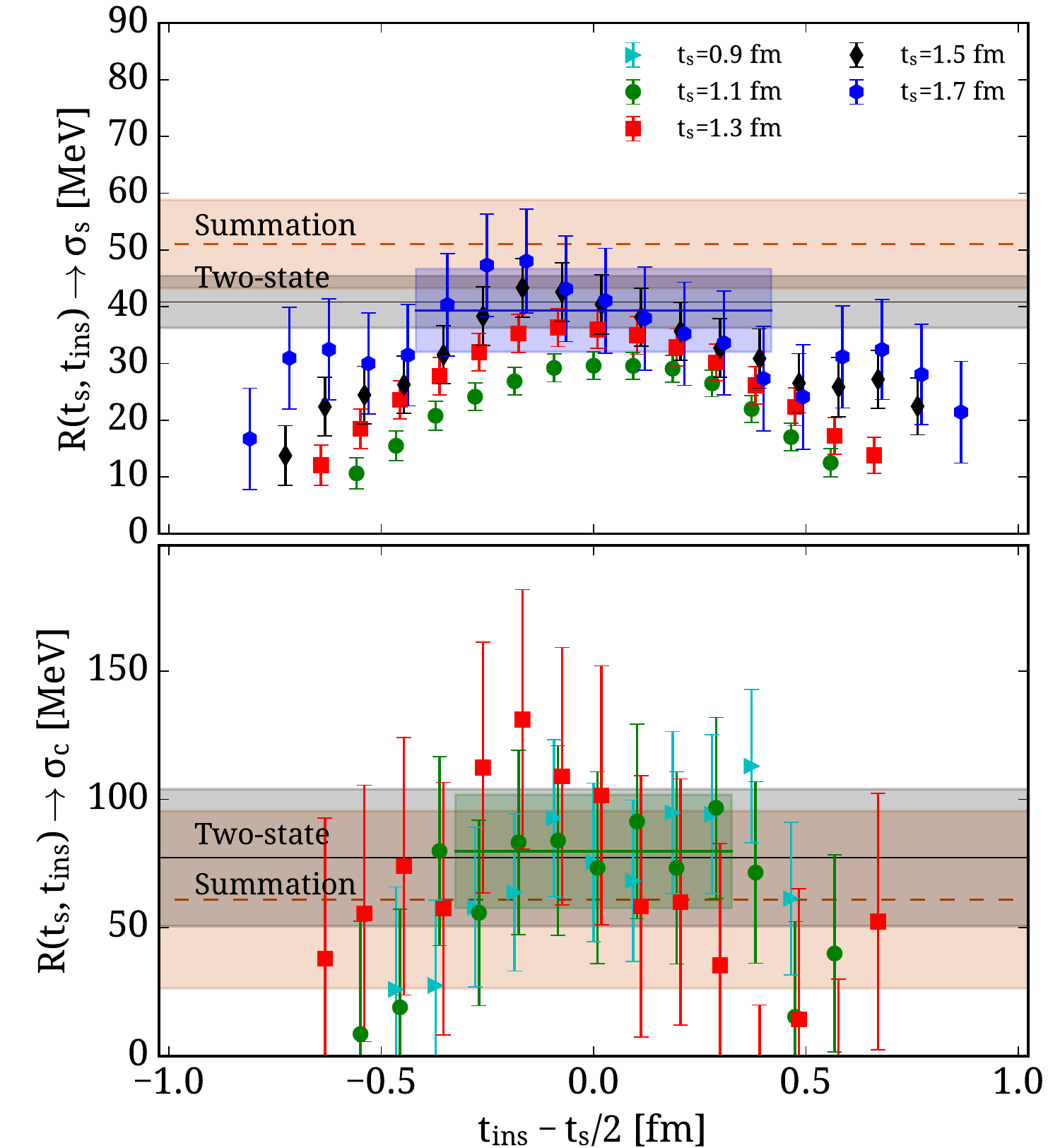}
  \caption{The ratio from which $\sigma_s$ (upper) and $\sigma_c$
    (lower) are extracted. The notation is the same as in
    Fig.~\ref{fig:plateau light}. }
  \label{fig:plateau strange}
\end{figure}
\begin{figure}[h!]
  \includegraphics[width=0.9\linewidth]{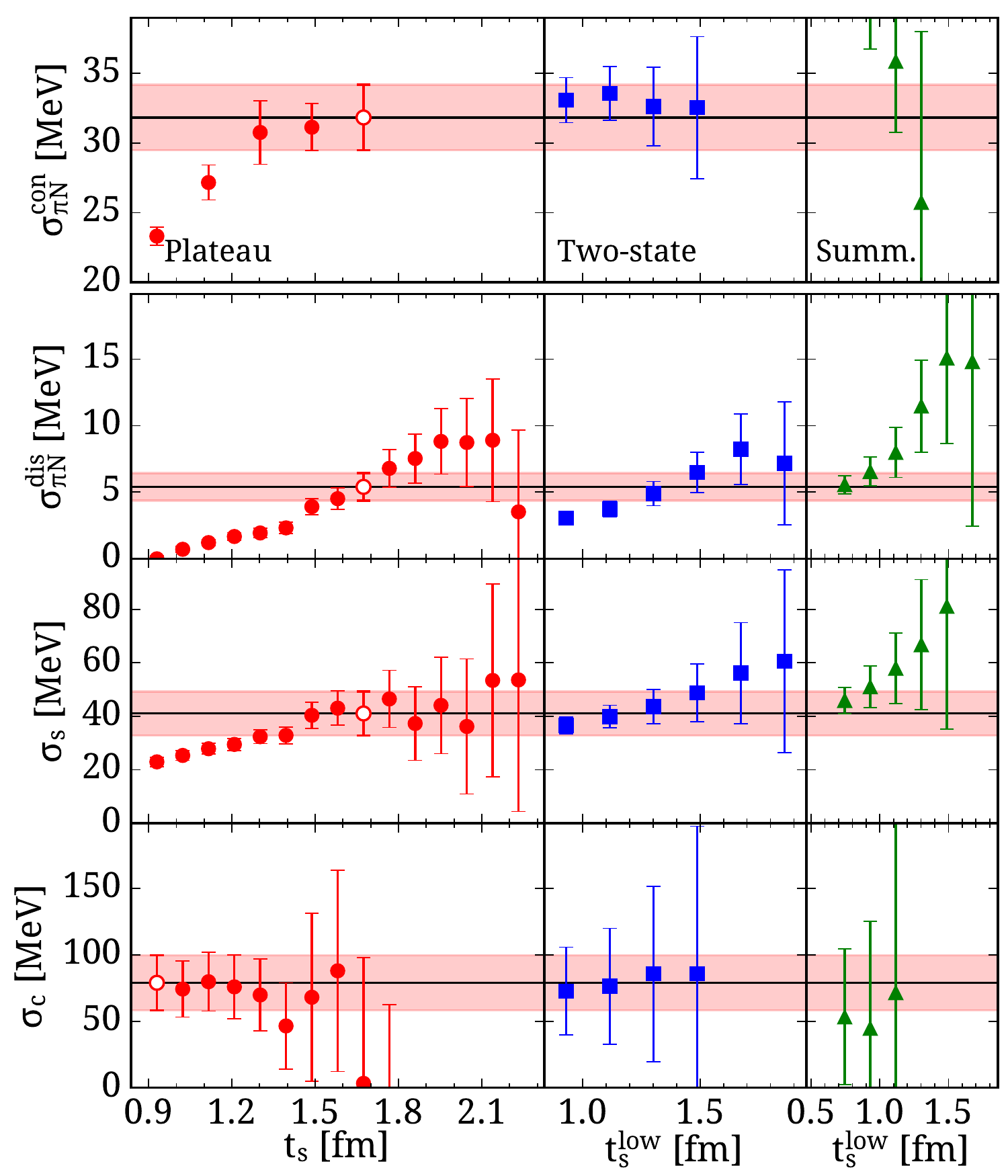}
  \caption{Results for $\sigma_f$ from the plateau (left column),
    two-state (center column) and summation (right column)
    methods. From top to bottom we show the results for the connected and
    disconnected contributions to $\sigma_{\pi N}$,  $\sigma_s$ and
    $\sigma_c$.  $t_s^{\rm low}$ is the smallest $t_s$ in the summation
    or two-state fits. The open symbol shows the selected final result
    and the red band its statistical error.}
  \label{fig:fits}
\end{figure}
\textit{Results:} In Figs.~\ref{fig:plateau light}
and~\ref{fig:plateau strange} we show results for the ratio of
Eq.~(\ref{eq:ratio}) for the light, the strange and the charm scalar operators,
respectively. As $t_s$ increases the ratios converge in the plateau
regions.  
We  fit to the
plateau when $t_s=1.7$~fm to extract results for $\sigma_{\pi N}$ and $\sigma_s$. For the charm sector, within the larger statistical errors, the plateaus are 
consistent already for $t_s=0.9$~fm. 
We observe that the disconnected contribution
to the $\sigma_{\pi N}$ is at the 10\% level, a value consistent with
the one found using an ensemble with pion mass of
373~MeV~\cite{Abdel-Rehim:2013wlz}.

The convergence of our results as we vary $t_s$ in the plateau method
is demonstrated in Fig.~\ref{fig:fits} where we also show the results
of the two-state fit and the summation method as we vary the lowest
 value of $t_s$ used in the fit, denoted by $t_s^{\rm low}$. Our final value is taken
from the plateau method after convergence with increasing $t_s$ is
demonstrated, as well as, when there is agreement with the two-state
fit. The summation method has typically larger errors but it
corroborates the plateau value. We quantify the systematic error due
to excited state effects by weighting all values extracted from the plateau with the fit
p-value for $t_s\ge$1.5~fm (0.9~fm) for the light and strange (charm)
$\sigma$-term and taking the variance.

Besides the systematic error due to the fit ranges, we have a
systematic error from the determination of the lattice spacing, which
we estimate by using the minimum and maximum values when different
physical quantities are used to set the
scale~\cite{Abdel-Rehim:2015pwa}. In addition, we  estimate the finite volume correction on $\sigma_{\pi N}$ by evaluating the
finite volume correction on the nucleon mass using baryon chiral
perturbation theory~\cite{AliKhan:2003ack}, in combination with the
Feynman-Hellmann theorem. 
We find a volume correction of
5\% on $\sigma_{\pi N}$, and  assume  this to be an upper bound also for $\sigma_{s}$ and
$\sigma_{c}$.

\begin{figure}[h]
  \includegraphics[width=0.9\linewidth]{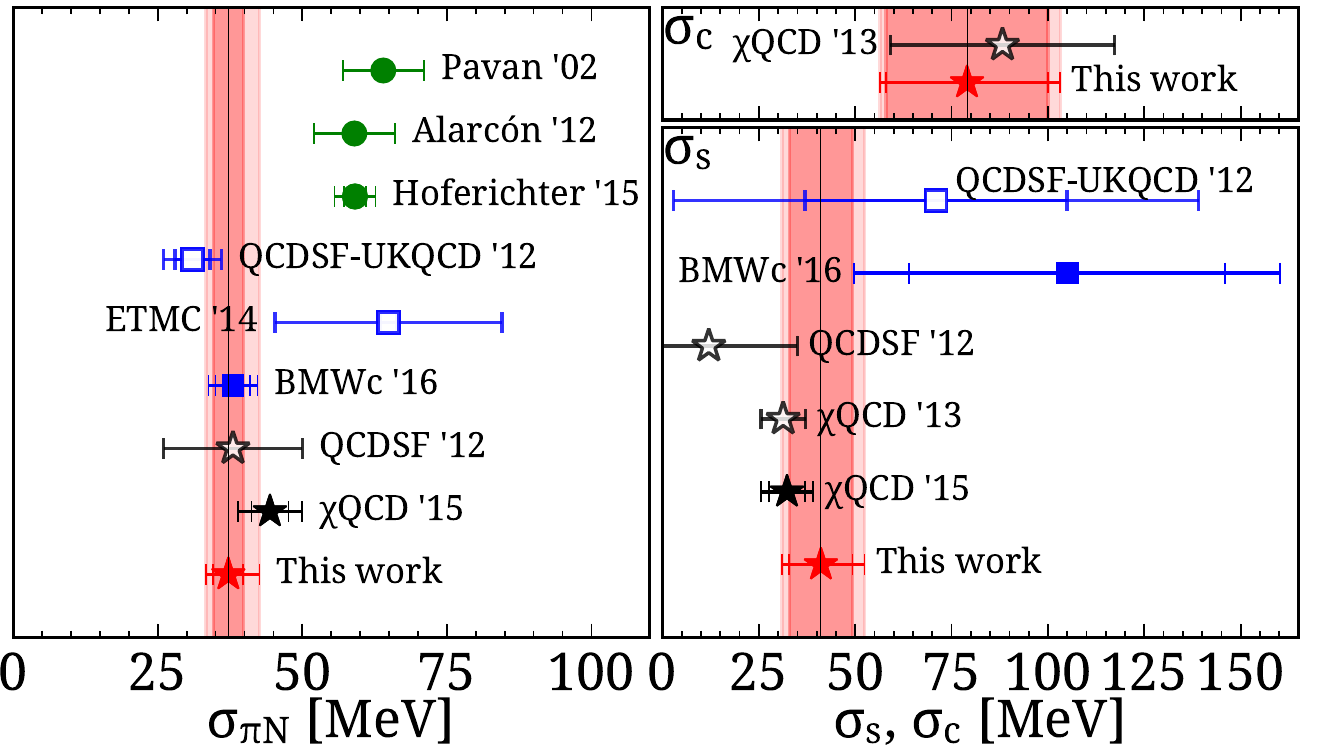}
  \caption{Comparison of recent results for the $\sigma$-terms  from phenomenology   (filled circles)~\cite{Pavan:2001wz,Alarcon:2011zs,Hoferichter:2012wf} and from lattice QCD shown with squares using  the Feynman-Hellmann theorem~\cite{Horsley:2011wr,Alexandrou:2014sha, Durr:2015dna} and  asterisks using the direct method~\cite{Bali:2011ks, Gong:2013vja, Yang:2015uis}. Open symbols use lattice QCD simulations that do not include the physical point. The smaller error bar is the statistical and the larger the total error.
}
  \label{fig:comparison}
\end{figure}


\textit{Conclusions}: Based on our analysis, 
the final values for $\sigma_{\pi N}$, $\sigma_s$ and $\sigma_c$ are:
\begin{align}
  \sigma_{\pi N} =& 37.2(2.6)_{\rm stat}(1.9)_{\rm exci}(1.0)_{\rm a}(^{1.8}_{0.0})_{\rm vol}~\textrm{MeV},&\nonumber\\
  \sigma_s = & 41.1(8.2)_{\rm stat}(4.7)_{\rm exci}(1.1)_{\rm a}(^{2.0}_{0.0})_{\rm vol}~\textrm{MeV}&\textrm{and} \nonumber\\
  \sigma_{c} = &79(21)_{\rm stat}(6)_{\rm exci}(2)_{\rm a}(^{4}_{0})_{\rm vol}~\textrm{MeV},&
\end{align}
where the first error is statistical and the rest are systematics due
to excited states (exci), finite lattice spacing (a), and
finite volume (vol) effects. The systematic errors are added linearly to
arrive at the errors in the final values given in the
abstract.   Alternatively, we provide our results in terms of the
dimensionless ratios, $f^N_f = \sigma_f/m_N$, obtaining:
\begin{align}
  f^N_{ud} &= 0.0399(28)(^{40}_{21}), &f^N_s = 0.0440(88)(^{72}_{51})\,\,\,\textrm{and}&\nonumber\\
  f^N_c &= 0.085(22)(^{11}_{\phantom{1}7})
\end{align}
using $m_N=933(8)$~MeV~\cite{Abdel-Rehim:2015owa}, where the first
error is statistical and the second is the sum of the systematic
uncertainties due to the excited states and  the finite
volume. The isovector matrix element $\langle N|\bar{u}u -
\bar{d}d|N\rangle$, has been computed~\cite{Abdel-Rehim:2015owa} and combining it with the isoscalar
matrix element we obtain the individual up- and down-quark
contributions for the proton and the neutron in the isospin limit via:
\begin{align}
  f_u^p = \frac{2 m_{ud} r}{r+1} \frac{\langle N|\bar{u}u|N\rangle}{m_N} & & f_u^n = \frac{2 m_{ud} r}{r+1} \frac{\langle N|\bar{d}d|N\rangle}{m_N} \nonumber\\
  f_d^p = \frac{2 m_{ud} }{r+1} \frac{\langle N|\bar{d}d|N\rangle}{m_N} & & f_d^n = \frac{2 m_{ud}}{r+1} \frac{\langle N|\bar{u}u|N\rangle}{m_N},
\end{align}
where 
$r=\frac{m_u}{m_d}=0.50(4)$ taken from
Ref.~\cite{Aoki:2013ldr}. 
We find
\begin{align}
  f_u^p = 0.0149(17)(^{21}_{14}),&               & f_u^n = 0.0117(15)(^{18}_{12}), \nonumber\\
  f_d^p = 0.0234(23)(^{27}_{16})&\,\textrm{and}\,& f_d^n = 0.0298(23)(^{30}_{16}),
\label{ratio values}
\end{align}
and for  the  $y_N$-parameter
\begin{align}
y_N \equiv \frac{2\langle N | \bar{s}s|N \rangle}{\langle N | \bar{u}u
  + \bar{d}d|N\rangle} = 0.075(16)(^{14}_{10}).
\end{align}
Isospin breaking can be estimated 
by comparing the values of $f_{u/d}^{p/n}$ in Eq.~(\ref{ratio values}) to those obtained when replacing the isovector matrix element with  the neutron-proton mass splitting $\Delta m_N|_{\rm QCD} = 2 m_{ud} \frac{1-r}{1+r}\langle N | \bar{u}u - \bar{d}d | N
\rangle$, and  using the value $\Delta m_N|_{\rm QCD} = 2.52(17)(24)$~MeV
  determined for
non-degenerate up- and down-quarks~\cite{Borsanyi:2014jba}.
We find a systematic error due to isospin breaking of maximum 1\% on $f_{u/d}^{p/n}$ i.e. an order of magnitude smaller than the other errors.

In Fig.~\ref{fig:comparison} we 
compare our values with recent results from phenomenology and lattice QCD omitting analyses that include   simulations with pion masses larger than 500~MeV.
As can be seen, our value for $\sigma_{\pi N}$ is in perfect agreement
with the most recent value determined using the Feynman-Hellmann theorem~\cite{Durr:2015dna}, which corroborates the consistency
between the two methods. Such a value of $\sigma_{\pi N}$ 
is consistent with a small SU(3) breaking in conjunction with a small
violation of the OZI rule. However, there is tension 
 with recent analyses based on the Roy-Steiner equations
and experimental data on pionic atoms. The lattice QCD value implies a
considerably smaller value of the $\pi N$-scattering lengths in
disagreement with pionic-atom phenomenology. A thorough investigation of the lattice systematics, as well as,   an evaluation of the $\pi
N$-scattering lengths within lattice QCD~\footnote{J. M. Alarc\'on, private communication.} can, thus,  provide crucial input in
resolving this puzzle.


\textit{Acknowledgments:} We thank all members of ETMC for their
cooperation.  This work used computational resources from the Swiss
National Supercomputing Centre (CSCS) under project ID s540 and s625,
from the John von Neumann-Institute for Computing on the Juropa system
and the BlueGene/Q system Juqueen at the research center in J\"ulich,
from the Gauss Centre for Supercomputing on HazelHen (HLRS) and PRACE
project access to the Tier-0 computing resources Curie (CEA), Fermi
(CINECA) and SuperMUC (LRZ).
K. H. and Ch. K. acknowledge support from the Cyprus Research Promotion
Foundation under contract T$\Pi$E/$\Pi\Lambda$HPO/0311(BIE)/09.

\bibliography{refs}
\end{document}